\documentclass[sigconf]{acmart}

\AtBeginDocument{%
  \providecommand\BibTeX{{%
    \normalfont B\kern-0.5em{\scshape i\kern-0.25em b}\kern-0.8em\TeX}}}

\setcopyright{acmcopyright}
\copyrightyear{2018}
\acmYear{2018}
\acmDOI{10.1145/1122445.1122456}

\acmConference[Woodstock '18]{Woodstock '18: ACM Symposium on Neural
  Gaze Detection}{June 03--05, 2018}{Woodstock, NY}
\acmBooktitle{Woodstock '18: ACM Symposium on Neural Gaze Detection,
  June 03--05, 2018, Woodstock, NY}
\acmPrice{15.00}
\acmISBN{978-1-4503-XXXX-X/18/06}

\usepackage{booktabs}
\usepackage{xac}

\usepackage{todonotes}

\begin{document}

\title{FaceOff: Detecting Face Touching with a Wrist-Worn Accelerometer}


\author{Xiang `Anthony' Chen}
\email{xac@ucla.edu}
\affiliation{UCLA HCI Research}

\renewcommand{\shortauthors}{Trovato and Tobin, et al.}

\begin{teaserfigure}
  \includegraphics[width=1\textwidth]{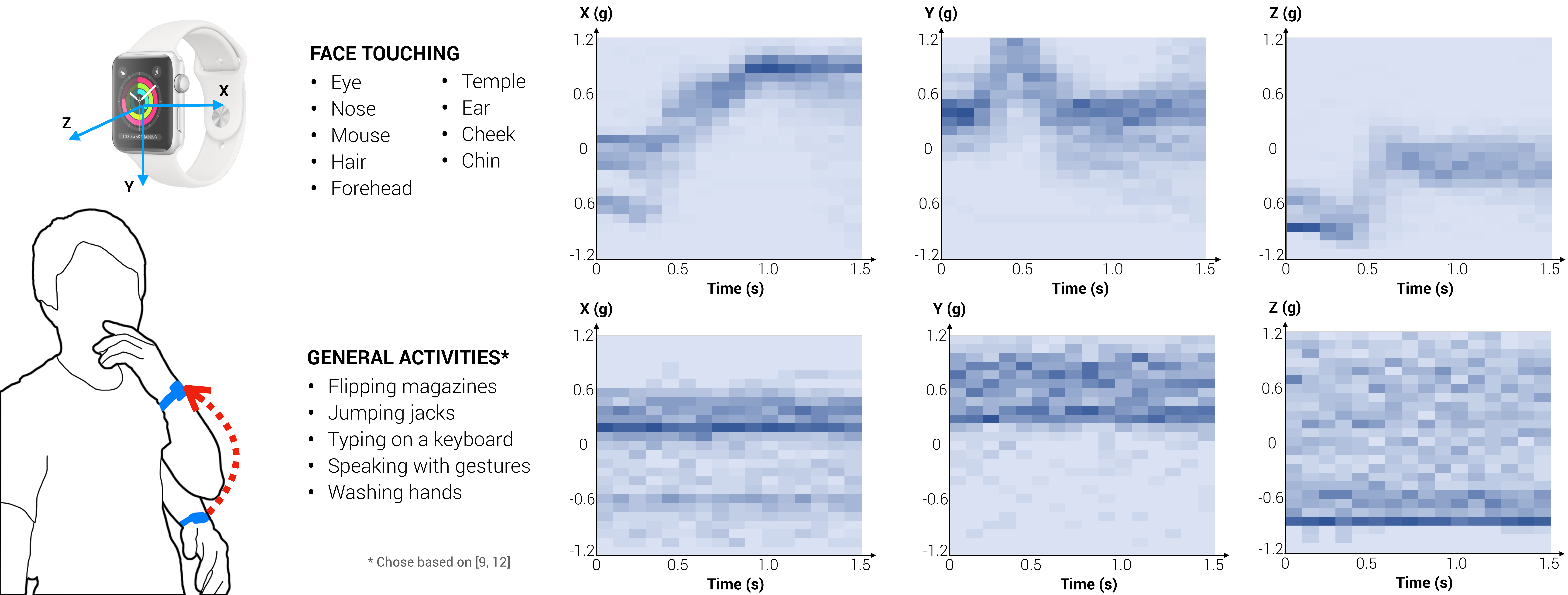}
  \vspace{-1.5em}
  \caption{We report evidence that demonstrates the potentials and limitations of using a commodity wrist-worn accelerometer to detect face-touching behavior based on the specific motion pattern of raising one's hand towards the face, detecting 82 out of 89 face touches with a false positive rate of 0.59\% in a preliminary study.}
  \label{fg:teaser}
\end{teaserfigure}

\begin{abstract}
  According to the CDC, one key step of preventing oneself from contracting coronavirus (COVID-19) is to avoid touching eyes, nose, and mouth with unwashed hands. However, touching one's face is a frequent and spontaneous behavior---one study observed subjects touching their faces on average 23 times per hour. Creative solutions have emerged amongst some recent commercial and hobbyists' projects, yet most either are closed-source or lack validation in performance. We develop FaceOff---a sensing technique using a commodity wrist-worn accelerometer to detect face-touching behavior based on the specific motion pattern of raising one's hand towards the face. We report a survey (N=20) that elicits different ways people touch their faces, an algorithm that temporally ensembles data-driven models to recognize when a face touching behavior occurs and results from a preliminary user testing (N=3 for a total of about 90 minutes).
\end{abstract}

\maketitle




\section{Introduction}

According to the Center of Disease Control and Prevention (CDC), one key step of preventing oneself from contracting coronavirus (COVID-19) is to avoid touching eyes, nose, and mouth with unwashed hands\footnote{\url{https://www.cdc.gov/coronavirus/2019-ncov/hcp/guidance-prevent-spread.html}}. Pathogens picked up by our hands can enter the throat and lungs through mucous membranes on the face.


However, touching one's face is a frequent and spontaneous behavior---Kwok \etal observed subjects touching their faces on average 23 times per hour \cite{kwok2015face}, where 44\% of the contacts were made with a mucous membrane.
To reduce face touching, creative solutions have emerged amongst some commercial and hobbyists' projects\footnote{\url{https://immutouch.com/}} \footnote{\url{https://blog.arduino.cc/2020/03/10/this-pair-of-arduino-glasses-stops-you-from-touching-your-face/}} \footnote{\url{https://www.media.mit.edu/projects/saving-face/overview/}}
since the outbreak of COVID-19. However, most are closed-source and/or lack validation in performance. 

We develop FaceOff---a sensing technique using a commodity wrist-worn accelerometer to detect face-touching behavior based on the specific motion pattern of raising one's hand towards the face (\fgref{teaser}).
We consider the touching of both mucosal and nonmucosal facial areas. Although it is touching the mucosal area that might cause infection, detecting touching both types of areas can more strongly raise people's awareness of avoiding touching their face at all. In a survey, we asked 20 participants\footnote{Demographic information reported in Appendix A} to describe where they would naturally touch their face; the result shows a wide range of facial areas (\fgref{survey}).

\fg{survey}{survey}{1.0}{Our preliminary survey (N=20) shows that people touch a wide range of facial parts. Our goal is to detect the touching of both mucosal and nonmucosal areas to strongly promote the awareness of avoiding touching one's face at all. \vspace{-2em}}

To detect face touching, we chose accelerometer---a common and low-cost sensor available in most wrist-worn devices (\eg smart watches, fitness trackers). The use of accelerations has shown promises in detecting body-tapping behavior without the need to instrument the body \cite{chen2016bootstrapping}, although detecting face touching has not been explored in prior work. We hypothesize that accelerometer can detect face touching by recognizing the hand's motion pattern---the unique course of accelerations and orientations as the hand moves towards the face. 
However, one limitation is that accelerometer cannot detect the actual contact with the face. For example, adjusting one's eyeglasses would appear highly similar to touching the eyes.

Based on the reported face touching,
we develop a data collection protocol. Due to the COVID-19 pandemic, we had limited access to participants for data collection. Thus we gathered training data only from the first author. We then develop an algorithm that temporally ensembles data-driven Random Forest classifiers to binarily detect whether a person touches their face. We analyze and validate this approach in a preliminary study on three other participants, each of which wore the device for 30 minutes with intermittent prompts to touch their face\footnote{Each participant thoroughly cleaned their hands and the physical objects they touched prior to the study} while conducting their own daily activities. As a result, 82 out of 89 (92\%) face touching actions were detected with a false positive rate of 0.59\%. 

{\bf Contributions} of this work are as follows.
\begin{itemize}
    \item To the best of our knowledge, the first reported evidence of the potentials and limitations of detecting face touching using a commodity wrist-worn accelerometer;
    \item A systematic protocol of data collection, training and testing, which can be adopted by future research to explore other sensing solutions for detecting face touching to combat the COVID-19 pandemic.
\end{itemize}







\section{Related Work}
There is a large body of work on wearable health sensing, for which we refer the readers to Pantelopoulos \etal's survey \cite{pantelopoulos2009survey}. Our review is focused on three prior research topics most related to our work.

{\bf Body-centric interaction} leverages (intentionally) tapping on different body parts, which can be used to trigger specific digital information or functions \cite{harrison2010skinput, chen2012extending}.
Similar examples include tapping on the body during running or cycling to control one's smart devices \cite{hamdan2017run,vechev2018movespace}, or moving a smartphone to the mouth for activating speech input \cite{yang2019proxitalk}.
Chen \etal demonstrate using an inertial measurement unit (IMU) and a phone's front camera to detect spatial interaction around a user's body \cite{chen2014around}, and later a single IMU alone to detect tapping the phone on a number of on-body locations \cite{chen2016bootstrapping}.
However, detecting face touching has never been investigated.

{\bf Detecting eating/drinking}, also involving a hand's motion relative to one's face, has been explored in prior work using a wrist-worn inertial sensor (\eg \cite{dong2013detecting, thomaz2015practical,thomaz2017exploring}).
However, detecting face touching presents new challenges: 
\one unlike eating, the person might no longer be constrained in a seated position; and
\two the motion pattern of one's hand is no longer limited to delivering food to the mouth but could vary across touching a range of facial parts.




{\bf Sensing personal hygienic behaviors} is related to the purpose of detecting face touching and in the literature is dominated by two specific activities: hand washing and tooth brushing.
For hand washing, various locations of wearable sensors have been explored: Li \etal use IMU on a wrist-worn device to detect whether a user's hand washing follows WHO's 13-step recommendation \cite{Li2018}.
Zhang \etal developed a ring device with fluid sensors for real-time monitoring of hand-hygiene compliance \cite{zhang2019smart}.
Kutafina \etal enabled hand hygiene training in medical education using a forearm Electromyography (EMG) sensor \cite{kutafina2015wearable}.
For tooth brushing, Hong \etal used an accelerometer + RFID sensor combination mounted on the back of a user's hand to recognize activities including tooth brushing
\cite{hong2008activity}.
Huang and Lin used a magnet attached to a manual toothbrush and an off-the-shelf smartwatch to detect fine-grained brushing behaviors on individual teeth \cite{huang2016toothbrushing}.
Wijayasingha and Lo proposed a wearable sensing framework using IMU for monitoring both hand washing and tooth brushing \cite{wijayasingha2016wearable}.
However, no prior work has addressed sensing the hygienic behavior of face touching.




\section{Data Collection}
We collected accelerometer data from the first author (male, 32, right-handed) wearing an Apple Watch 2 (100 Hz sampling rate) on the left wrist.


\subsection{Behavioral Variations}
The independent variable is {\it Behavior} (Touch \vs No touch).

%
For Touch, we consider where and how.
For where to touch, we include various parts of the face based on CDC's guidelines (eyes, nose, and mouth) as well as other frequently touched areas indicated in the aforementioned survey: hair, forehead, temple, ear, cheek, and chin.
Since our face is symmetric, we also consider left \vs right side for certain facial parts (\eg left \vs right ear).
In terms of how to touch, we cover both transient touch (\eg a quick scratch on the nose) and lingering touch (\eg holding and rubbing the chin).
We also consider hand placement (before face-touching), following and extending the experiment design in \cite{chen2016bootstrapping}. Specifically, we consider high, middle and low hand placement:
for high placement the user sits and places the forearms on a desk in front of them;
for middle placement the user sits and places the hands on the laps;
for low placement the user stands and lowers their arms naturally around the waist.
Different initial hand placements result in different trajectories the hand travels towards the face.

For No Touch, we adopt the set of general daily activities used in \cite{laput2016viband, kang2019minuet}: flipping magazines, jumping-jacks, typing on a keyboard, speaking while gesturing and washing hands and washing hands.


\subsection{Procedure}
The data collection was split into three sessions corresponding to the three hand placement conditions for Touch. In between sessions the user took off and then put the watch back on, which accounted for the possible variance of how the watch was worn (\eg position, orientation, tightness).

At the beginning of a session, the user thoroughly cleaned the watch and washed and dried the hands.
We then started with collecting data for Touch where the user was prompted with instructions on the watch to touch a facial part either transiently or lingeringly. 
The user would tap anywhere on the watch screen, which started the data collection of a facing touching trial. The user then proceeded to touch the designated facial part. The data collection ended automatically after our empirically pre-defined 1.5$s$ window (which covers the time taken to raise one's hand and engage it in the touching of a facial part).
Each facial part was touched eight times. For symmetric parts (\eg eyes, ears) these trials were evenly split between left/right sides. The order of facial parts was randomized to avoid temporal dependence of behavior. 
Further, across all the trials, how to touch was evenly and randomly split between transient \vs lingering. For example, the user would be prompted to ``touch left cheek lingeringly''.

Next, we collected data for No Touch where
the user simply performed each aforementioned activity for 30$s$ while the system randomly collected 10 samples.
In total, we collected

1 user $\times$ 3 sessions per user $\times$\\
per session:
    (8 trials per facial part $\times$ 9 facial parts +\\
    10 trials per general activity $\times$ 5 general activities) $\times$\\
= {\bf 366} data points

\section{Detecting Face Touching}
\subsection{Visualization \& Featurization}
To visualize the data, we first downsample each trial to 15 bins, each containing a time interval of $0.1s$ of accelerometer data. 
\fgref{teaser} shows an aggregated view of the aforementioned collected data: each column in each chart corresponds to one bin that contains a distribution of accelerometer readings within the bin's time interval.
We can see how face touching and general activities exhibit distinct motion patterns across all the axes (note that the axis alignment is configured for wearing the watch on the left wrist).
For example, despite touching different facial areas, the X axis almost always points up, as shown in the latter half of the time window. 
For the Y axis, the peak at the beginning of the time window is likely caused by the motion of rotating one's forearm around the elbow and towards the face, where Y would first point up and then `flattens' when the hand is raised to the face level.


Rather than individual bins, our featurization focuses on characterizing the entire 1.5$s$ window of data as a distribution. Specifically, our features consist of
the statistical summary (sum, mean, median, standard deviation, coefficient of variation, zero crossing, mean/median absolute deviation) and shape-related measurements (skewness and kurtosis). In total, we use these 10 features per axis $\times$ 3 axes = 30 features.

\subsection{Temporal Ensemble of Data Driven Models}
Since the goal of detecting face touching is prevention, it is beneficial to detect a facing touching action before its completion. In other words, the question is whether we can use a smaller time window than 1.5$s$. 
To investigate this possibility, we train a series of Random Forest\footnote{
    Implementation details about hyperparameter tuning is reported in the Appendix B.
} models based on the collected data from the one user. 
Each model uses only a partial window of a trial's data for inference, \eg at $t=1.0s$, a model $f_{1.0s}$ only uses data up to that point, \ie between 0 and $1.0s$. 
We compute the F1 scores from a 10-fold cross validation. 

\fgh{onsets}{onsets}{1.0}{F1 scores of using a partial time window of data (\eg for $t=1.0s$, a model only uses data from 0 to 1.0$s$) to detect face touching.
\vspace{-2em}
}

As shown in \fgref{onsets}, as we `wait for' more data, the F1 score expectedly increases and beyond $1.0s$ the F1 score seems to flatten.
Thus we select $1.0s$ as the starting point from which a model would detect face touching.
Specifically, we train different models to processes different incoming data at $T = \{1.0s, 1.1s, 1.2s, 1.3s, 1.4s, 1.5s\}$.
At $t \in T$, the current data is $\mathbf{x}_t$ and the corresponding model is $f_t: \mathbf{x} \rightarrow \{-1, 1\}$ (1 for face touching and -1 for no face touching).
A final result is obtained via voting where each model's vote is its result weighted by the F1 score:
$\text{sgn}(\sum_{t \in T}w_t f_t(\mathbf{x}_t))$, 
where 
$\text{sgn}$ is the sign function,
$w_t=\exp(\lambda(\text{F1}_t - \min_T{\text{F1}}))$
and $\lambda$ is constant to scale the difference amongst the F1 scores (we use $\lambda=10$).



We tested this approach of temporally ensembling data-driven models in a preliminary setting, as described below.

\section{Preliminary Testing}
We recruited another three\footnote{Due to the COVID-19 pandemic, we had to recruit the only three people living in the same household as the first author} participants (male, 18; male, 22; female 28; all right-handed). 
Two participants reported touching their face at least once an hour wheres the other was unaware of the frequency. 
Each participant identified five most common ways they touched their face: all three reported ear and nose; eye, cheek, chin and forehead were reported twice; only one participant reported hair.


\subsection{Task \& Procedure}
Participants were asked to wear the watch on the left wrist (\ie the non-dominant hand) for a period of about 30 minutes while performing daily routine activities of their choice.
P1 sat on a reclined chair and browse her phone, P2 stood and walked around the house while listening to lectures on the phone via headphones, and P3 sat down and worked on programming tasks on a laptop computer. 
Admittedly, this protocol only captured a brief snapshot of how well the system might work in the context of a specific user-chosen daily activity; we consider a fully in-the-wild study with more participants as future work (after the pandemic).

Before the period started, each participant thoroughly washed/dried their hands and cleaned the watch and objects they expected to touch in the next 30 minutes.
During the period, each participant was randomly prompted 30 times (on average once per minute) to use their left hand to touch their face on the five facial parts they identified earlier. 
Each prompt consisted of two steps.
Firstly the watch interrupted the participant's activity with vibration that prompted them to raise the watch and read the instruction of touching a specific facial part.
If a participant were to touch their face immediately, the motion would be unnatural as it would artificially start from a wrist-raising posture.
Instead, we asked the participant to press the `confirm' button, and resumed their activity; then in four seconds, the watch vibrated again, following which the participant would now touch their face at the specified part.

\begin{table}[t]

    \begin{tabular}{lcccc}
    \toprule
                                 & P1    & P2    & P3    & Overall         \\
    \midrule
    Face touching detected       & 25/29 & 27/30 & 30/30 & {\bf 82}/89 \\
    False positives rate & 0.66\%  & 0.57\%  & 0.55\%  & 0.59\%            \\
    \bottomrule  
    \end{tabular}
    \caption{Preliminary testing results of ensembling models using data at \{1.0s, 1.1s, 1.2s, 1.3s, 1.4s, 1.5s\}.}
    \label{tb:results1}



    \begin{tabular}{lcccc}
    \toprule
                                & P1    & P2    & P3    & Overall         \\
    \midrule
    Face touching detected       & 23/29 & 27/30 & 30/30 & 80/89 \\
    False positives rate & 0.62\%  & 0.56\%  & 0.50\%  & {\bf 0.56\%}         \\
    \bottomrule  
    \end{tabular}
    \caption{Preliminary testing results of using only the entire 1.5$s$ of data at the end of the time window.}
    \label{tb:results2}
    \vspace{-2.5em}
\end{table}

\subsection{Data Logging \& Labeling}

We used the same 1.5$s$ as the sliding window length at a rate of four FPS (\ie a 83.33\% overlap between subsequent frames).
Face touching was labeled on the three seconds of data after the second vibration that prompted the participant to actually touch their face. The rest of the data was labeled as no face touching.

\subsection{Results}



We analyzed the logged data offline using models developed earlier from a different user than the three participants. The last three minutes of P1's data was lost due to technical issues, which lost one face touching trial and some no touching data. 



As shown in Table~\ref{tb:results1}, our ensemble approach detected 82 out of 89 face touching actions. 
We found that sometimes we did not have to wait until the end to finalize the vote result---a majority vote could be formed at $t<1.5s$.
Specifically, amongst the 82 face touches we found, 18 were detected at $t=1.3s$, 28 were at $t=1.4s$ and the rest at $1.5s$.
In comparison, if we were to use only the entire 1.5$s$ of data at the end of the time window, the recall rate would be slightly lower (80/89), as shown in Table~\ref{tb:results2}. 

We also compute the false positive rate, which is the percentage of no-face-touching instances labeled as face-touching.
As shown in Table \ref{tb:results1} \& \ref{tb:results2}, the two approaches achieved similar false positive rates (with only a 0.03\% difference).


Overall, results show that it is feasible to detect face touching using a wrist-worn accelerometer: our ensemble approach achieved slightly higher recall rate without compromising the prevention of false positives. Meanwhile, for over half the time (18+28), an earlier majority vote was able to preemptively determine a face touching action before collecting the entire window of accelerometer data.


\section{Limitations \& Future Work}
{\bf The inability to detect actual hand-face contact}.
Based on our observations, a number of false positives are caused by behavior that resembles touching one's face, \eg scratching the back of one's neck, raising eyeglasses, picking up a phone call, drinking water. 
However, it is a reasonable design choice to alert a user of such false positives, as their hand is still brought to close proximity to the face even without touching. 
It would be appropriately aggressive to promote an awareness of trying to keep one's hands off the face as much as possible regardless of whether actual contact is made.

{\bf The current set of no-touch activities needs to be expanded}: 
as future work explores other types of sensors, \eg  proximity detection between the hand and the face, it is also possible to use camera instrumented in the environment, using rich visual information to distinguish face touching from certain near-miss activities (\eg eating, drinking). Using cameras, it is also possible to annotate naturally-occurring face-touching behaviors, which addresses the currently-controlled experiment task setting. However, privacy issues and line-of-sight constraints are two long-standing challenges for camera-based solutions.

{\bf One-handed detection only} is another apparent limitation. While it is uncommon to ask users to wear two watches, future work can explore an alternate form factor, \eg wrist-band accelerometer (\eg Fitbit-like devices), which would be more socially and economically appropriate to wear on both hands.

{\bf Feedback mechanisms to effect behavior change} would be an important next-step now that we have demonstrated a proof-of-concept mechanism for detecting face touching.
Future work should investigate both visual and vibrotactile feedbacks that alert a person both at the onset of face touching for prevention and after one touches the face for feedback that encourages behavior change. Longitudinal study should be conducted to track people's behavior change given the detection and alerts of face touching.

{\bf Larger-scale data collection and user testing} should be conducted (when available) to account for the possibly different ways people touch their face, to improve the data-driven models and to obtain statistically-significant performance evaluation results.

\bibliographystyle{ACM-Reference-Format}

\bibliography{references,general,mypubs}


\begin{thebibliography}{20}


\ifx \showCODEN    \undefined \def \showCODEN     #1{\unskip}     \fi
\ifx \showDOI      \undefined \def \showDOI       #1{#1}\fi
\ifx \showISBNx    \undefined \def \showISBNx     #1{\unskip}     \fi
\ifx \showISBNxiii \undefined \def \showISBNxiii  #1{\unskip}     \fi
\ifx \showISSN     \undefined \def \showISSN      #1{\unskip}     \fi
\ifx \showLCCN     \undefined \def \showLCCN      #1{\unskip}     \fi
\ifx \shownote     \undefined \def \shownote      #1{#1}          \fi
\ifx \showarticletitle \undefined \def \showarticletitle #1{#1}   \fi
\ifx \showURL      \undefined \def \showURL       {\relax}        \fi
\providecommand\bibfield[2]{#2}
\providecommand\bibinfo[2]{#2}
\providecommand\natexlab[1]{#1}
\providecommand\showeprint[2][]{arXiv:#2}

\bibitem[\protect\citeauthoryear{Chen and Li}{Chen and Li}{2016}]%
        {chen2016bootstrapping}
\bibfield{author}{\bibinfo{person}{Xiang~`Anthony' Chen} {and}
  \bibinfo{person}{Yang Li}.} \bibinfo{year}{2016}\natexlab{}.
\newblock \showarticletitle{Bootstrapping User-Defined Body Tapping Recognition
  with Offline-Learned Probabilistic Representation}. In
  \bibinfo{booktitle}{\emph{Proceedings of the 29th Annual Symposium on User
  Interface Software and Technology}}. ACM, \bibinfo{pages}{359--364}.
\newblock


\bibitem[\protect\citeauthoryear{Chen, Marquardt, Tang, Boring, and
  Greenberg}{Chen et~al\mbox{.}}{2012}]%
        {chen2012extending}
\bibfield{author}{\bibinfo{person}{Xiang~`Anthony' Chen},
  \bibinfo{person}{Nicolai Marquardt}, \bibinfo{person}{Anthony Tang},
  \bibinfo{person}{Sebastian Boring}, {and} \bibinfo{person}{Saul Greenberg}.}
  \bibinfo{year}{2012}\natexlab{}.
\newblock \showarticletitle{Extending a mobile device's interaction space
  through body-centric interaction}. In \bibinfo{booktitle}{\emph{Proceedings
  of the 14th international conference on Human-computer interaction with
  mobile devices and services}}. ACM, \bibinfo{pages}{151--160}.
\newblock


\bibitem[\protect\citeauthoryear{Chen, Schwarz, Harrison, Mankoff, and
  Hudson}{Chen et~al\mbox{.}}{2014}]%
        {chen2014around}
\bibfield{author}{\bibinfo{person}{Xiang~`Anthony' Chen},
  \bibinfo{person}{Julia Schwarz}, \bibinfo{person}{Chris Harrison},
  \bibinfo{person}{Jennifer Mankoff}, {and} \bibinfo{person}{Scott Hudson}.}
  \bibinfo{year}{2014}\natexlab{}.
\newblock \showarticletitle{Around-body interaction: sensing and interaction
  techniques for proprioception-enhanced input with mobile devices}. In
  \bibinfo{booktitle}{\emph{Proceedings of the 16th international conference on
  Human-computer interaction with mobile devices and services}}. ACM,
  \bibinfo{pages}{287--290}.
\newblock


\bibitem[\protect\citeauthoryear{Dong, Scisco, Wilson, Muth, and Hoover}{Dong
  et~al\mbox{.}}{2013}]%
        {dong2013detecting}
\bibfield{author}{\bibinfo{person}{Yujie Dong}, \bibinfo{person}{Jenna Scisco},
  \bibinfo{person}{Mike Wilson}, \bibinfo{person}{Eric Muth}, {and}
  \bibinfo{person}{Adam Hoover}.} \bibinfo{year}{2013}\natexlab{}.
\newblock \showarticletitle{Detecting periods of eating during free-living by
  tracking wrist motion}.
\newblock \bibinfo{journal}{\emph{IEEE journal of biomedical and health
  informatics}} \bibinfo{volume}{18}, \bibinfo{number}{4}
  (\bibinfo{year}{2013}), \bibinfo{pages}{1253--1260}.
\newblock


\bibitem[\protect\citeauthoryear{Hamdan, Kosuru, Corsten, and Borchers}{Hamdan
  et~al\mbox{.}}{2017}]%
        {hamdan2017run}
\bibfield{author}{\bibinfo{person}{Nur Al-huda Hamdan},
  \bibinfo{person}{Ravi~Kanth Kosuru}, \bibinfo{person}{Christian Corsten},
  {and} \bibinfo{person}{Jan Borchers}.} \bibinfo{year}{2017}\natexlab{}.
\newblock \showarticletitle{Run\&Tap: Investigation of On-Body Tapping for
  Runners}. In \bibinfo{booktitle}{\emph{Proceedings of the 2017 ACM
  International Conference on Interactive Surfaces and Spaces}}.
  \bibinfo{pages}{280--286}.
\newblock


\bibitem[\protect\citeauthoryear{Harrison, Tan, and Morris}{Harrison
  et~al\mbox{.}}{2010}]%
        {harrison2010skinput}
\bibfield{author}{\bibinfo{person}{Chris Harrison}, \bibinfo{person}{Desney
  Tan}, {and} \bibinfo{person}{Dan Morris}.} \bibinfo{year}{2010}\natexlab{}.
\newblock \showarticletitle{Skinput: appropriating the body as an input
  surface}. In \bibinfo{booktitle}{\emph{Proceedings of the SIGCHI conference
  on human factors in computing systems}}. \bibinfo{pages}{453--462}.
\newblock


\bibitem[\protect\citeauthoryear{Hong, Kim, Ahn, and Kim}{Hong
  et~al\mbox{.}}{2008}]%
        {hong2008activity}
\bibfield{author}{\bibinfo{person}{Yu-Jin Hong}, \bibinfo{person}{Ig-Jae Kim},
  \bibinfo{person}{Sang~Chul Ahn}, {and} \bibinfo{person}{Hyoung-Gon Kim}.}
  \bibinfo{year}{2008}\natexlab{}.
\newblock \showarticletitle{Activity recognition using wearable sensors for
  elder care}. In \bibinfo{booktitle}{\emph{2008 Second International
  Conference on Future Generation Communication and Networking}},
  Vol.~\bibinfo{volume}{2}. IEEE, \bibinfo{pages}{302--305}.
\newblock


\bibitem[\protect\citeauthoryear{Huang and Lin}{Huang and Lin}{2016}]%
        {huang2016toothbrushing}
\bibfield{author}{\bibinfo{person}{Hua Huang} {and} \bibinfo{person}{Shan
  Lin}.} \bibinfo{year}{2016}\natexlab{}.
\newblock \showarticletitle{Toothbrushing monitoring using wrist watch}. In
  \bibinfo{booktitle}{\emph{Proceedings of the 14th ACM Conference on Embedded
  Network Sensor Systems CD-ROM}}. \bibinfo{pages}{202--215}.
\newblock


\bibitem[\protect\citeauthoryear{Kang, Guo, Laput, Li, and Chen}{Kang
  et~al\mbox{.}}{2019}]%
        {kang2019minuet}
\bibfield{author}{\bibinfo{person}{Runchang Kang}, \bibinfo{person}{Anhong
  Guo}, \bibinfo{person}{Gierad Laput}, \bibinfo{person}{Yang Li}, {and}
  \bibinfo{person}{Xiang~'Anthony' Chen}.} \bibinfo{year}{2019}\natexlab{}.
\newblock \showarticletitle{Minuet: Multimodal Interaction with an Internet of
  Things}. In \bibinfo{booktitle}{\emph{To Appear at the ACM symposium on
  Spatial user interaction}}. ACM.
\newblock


\bibitem[\protect\citeauthoryear{Kutafina, Laukamp, and Jonas}{Kutafina
  et~al\mbox{.}}{2015}]%
        {kutafina2015wearable}
\bibfield{author}{\bibinfo{person}{Ekaterina Kutafina}, \bibinfo{person}{David
  Laukamp}, {and} \bibinfo{person}{Stephan~M Jonas}.}
  \bibinfo{year}{2015}\natexlab{}.
\newblock \showarticletitle{Wearable Sensors in Medical Education: Supporting
  Hand Hygiene Training with a Forearm EMG.}. In
  \bibinfo{booktitle}{\emph{pHealth}}. \bibinfo{pages}{286--291}.
\newblock


\bibitem[\protect\citeauthoryear{Kwok, Gralton, and McLaws}{Kwok
  et~al\mbox{.}}{2015}]%
        {kwok2015face}
\bibfield{author}{\bibinfo{person}{Yen Lee~Angela Kwok}, \bibinfo{person}{Jan
  Gralton}, {and} \bibinfo{person}{Mary-Louise McLaws}.}
  \bibinfo{year}{2015}\natexlab{}.
\newblock \showarticletitle{Face touching: A frequent habit that has
  implications for hand hygiene}.
\newblock \bibinfo{journal}{\emph{American journal of infection control}}
  \bibinfo{volume}{43}, \bibinfo{number}{2} (\bibinfo{year}{2015}),
  \bibinfo{pages}{112--114}.
\newblock


\bibitem[\protect\citeauthoryear{Laput, Xiao, and Harrison}{Laput
  et~al\mbox{.}}{2016}]%
        {laput2016viband}
\bibfield{author}{\bibinfo{person}{Gierad Laput}, \bibinfo{person}{Robert
  Xiao}, {and} \bibinfo{person}{Chris Harrison}.}
  \bibinfo{year}{2016}\natexlab{}.
\newblock \showarticletitle{Viband: High-fidelity bio-acoustic sensing using
  commodity smartwatch accelerometers}. In
  \bibinfo{booktitle}{\emph{Proceedings of the 29th Annual Symposium on User
  Interface Software and Technology}}. \bibinfo{pages}{321--333}.
\newblock


\bibitem[\protect\citeauthoryear{Li, Chawla, Li, Jain, Abowd, Starner, Zhang,
  and Pl{\"{o}}tz}{Li et~al\mbox{.}}{2018}]%
        {Li2018}
\bibfield{author}{\bibinfo{person}{Hong Li}, \bibinfo{person}{Shishir Chawla},
  \bibinfo{person}{Richard Li}, \bibinfo{person}{Sumeet Jain},
  \bibinfo{person}{Gregory~D. Abowd}, \bibinfo{person}{Thad Starner},
  \bibinfo{person}{Cheng Zhang}, {and} \bibinfo{person}{Thomas Pl{\"{o}}tz}.}
  \bibinfo{year}{2018}\natexlab{}.
\newblock \showarticletitle{{Wristwash}}. In
  \bibinfo{booktitle}{\emph{Proceedings of the 2018 ACM International Symposium
  on Wearable Computers - ISWC '18}}. \bibinfo{publisher}{ACM Press},
  \bibinfo{address}{New York, New York, USA}, \bibinfo{pages}{132--139}.
\newblock
\showISBNx{9781450359672}
\showISSN{15504816}
\urldef\tempurl%
\url{https://doi.org/10.1145/3267242.3267247}
\showDOI{\tempurl}


\bibitem[\protect\citeauthoryear{Pantelopoulos and Bourbakis}{Pantelopoulos and
  Bourbakis}{2009}]%
        {pantelopoulos2009survey}
\bibfield{author}{\bibinfo{person}{Alexandros Pantelopoulos} {and}
  \bibinfo{person}{Nikolaos~G Bourbakis}.} \bibinfo{year}{2009}\natexlab{}.
\newblock \showarticletitle{A survey on wearable sensor-based systems for
  health monitoring and prognosis}.
\newblock \bibinfo{journal}{\emph{IEEE Transactions on Systems, Man, and
  Cybernetics, Part C (Applications and Reviews)}} \bibinfo{volume}{40},
  \bibinfo{number}{1} (\bibinfo{year}{2009}), \bibinfo{pages}{1--12}.
\newblock


\bibitem[\protect\citeauthoryear{Thomaz, Bedri, Prioleau, Essa, and
  Abowd}{Thomaz et~al\mbox{.}}{2017}]%
        {thomaz2017exploring}
\bibfield{author}{\bibinfo{person}{Edison Thomaz}, \bibinfo{person}{Abdelkareem
  Bedri}, \bibinfo{person}{Temiloluwa Prioleau}, \bibinfo{person}{Irfan Essa},
  {and} \bibinfo{person}{Gregory~D Abowd}.} \bibinfo{year}{2017}\natexlab{}.
\newblock \showarticletitle{Exploring symmetric and asymmetric bimanual eating
  detection with inertial sensors on the wrist}. In
  \bibinfo{booktitle}{\emph{Proceedings of the 1st Workshop on Digital
  Biomarkers}}. \bibinfo{pages}{21--26}.
\newblock


\bibitem[\protect\citeauthoryear{Thomaz, Essa, and Abowd}{Thomaz
  et~al\mbox{.}}{2015}]%
        {thomaz2015practical}
\bibfield{author}{\bibinfo{person}{Edison Thomaz}, \bibinfo{person}{Irfan
  Essa}, {and} \bibinfo{person}{Gregory~D Abowd}.}
  \bibinfo{year}{2015}\natexlab{}.
\newblock \showarticletitle{A practical approach for recognizing eating moments
  with wrist-mounted inertial sensing}. In
  \bibinfo{booktitle}{\emph{Proceedings of the 2015 ACM International Joint
  Conference on Pervasive and Ubiquitous Computing}}.
  \bibinfo{pages}{1029--1040}.
\newblock


\bibitem[\protect\citeauthoryear{Vechev, Dancu, Perrault, Roy, Fjeld, and
  Zhao}{Vechev et~al\mbox{.}}{2018}]%
        {vechev2018movespace}
\bibfield{author}{\bibinfo{person}{Velko Vechev}, \bibinfo{person}{Alexandru
  Dancu}, \bibinfo{person}{Simon~T Perrault}, \bibinfo{person}{Quentin Roy},
  \bibinfo{person}{Morten Fjeld}, {and} \bibinfo{person}{Shengdong Zhao}.}
  \bibinfo{year}{2018}\natexlab{}.
\newblock \showarticletitle{Movespace: on-body athletic interaction for running
  and cycling}. In \bibinfo{booktitle}{\emph{Proceedings of the 2018
  international conference on advanced visual interfaces}}.
  \bibinfo{pages}{1--9}.
\newblock


\bibitem[\protect\citeauthoryear{Wijayasingha and Lo}{Wijayasingha and
  Lo}{2016}]%
        {wijayasingha2016wearable}
\bibfield{author}{\bibinfo{person}{Lahiru~NS Wijayasingha} {and}
  \bibinfo{person}{Benny Lo}.} \bibinfo{year}{2016}\natexlab{}.
\newblock \showarticletitle{A wearable sensing framework for improving personal
  and oral hygiene for people with developmental disabilities}. In
  \bibinfo{booktitle}{\emph{2016 IEEE Wireless Health (WH)}}. IEEE,
  \bibinfo{pages}{1--7}.
\newblock


\bibitem[\protect\citeauthoryear{Yang, Yu, Zheng, and Shi}{Yang
  et~al\mbox{.}}{2019}]%
        {yang2019proxitalk}
\bibfield{author}{\bibinfo{person}{Zhican Yang}, \bibinfo{person}{Chun Yu},
  \bibinfo{person}{Fengshi Zheng}, {and} \bibinfo{person}{Yuanchun Shi}.}
  \bibinfo{year}{2019}\natexlab{}.
\newblock \showarticletitle{ProxiTalk: Activate Speech Input by Bringing
  Smartphone to the Mouth}.
\newblock \bibinfo{journal}{\emph{Proceedings of the ACM on Interactive,
  Mobile, Wearable and Ubiquitous Technologies}} \bibinfo{volume}{3},
  \bibinfo{number}{3} (\bibinfo{year}{2019}), \bibinfo{pages}{1--25}.
\newblock


\bibitem[\protect\citeauthoryear{Zhang, Kadimisetty, Yin, Ruiz, Mauk, and
  Liu}{Zhang et~al\mbox{.}}{2019}]%
        {zhang2019smart}
\bibfield{author}{\bibinfo{person}{Xin Zhang}, \bibinfo{person}{Karteek
  Kadimisetty}, \bibinfo{person}{Kun Yin}, \bibinfo{person}{Carlos Ruiz},
  \bibinfo{person}{Michael~G Mauk}, {and} \bibinfo{person}{Changchun Liu}.}
  \bibinfo{year}{2019}\natexlab{}.
\newblock \showarticletitle{Smart ring: a wearable device for hand hygiene
  compliance monitoring at the point-of-need}.
\newblock \bibinfo{journal}{\emph{Microsystem Technologies}}
  \bibinfo{volume}{25}, \bibinfo{number}{8} (\bibinfo{year}{2019}),
  \bibinfo{pages}{3105--3110}.
\newblock


\end{thebibliography}

\appendix

\section{Survey Participants Information}
We disseminated a face touching survey via an business communication platform amongst members of our research group and received 20 responses.
The participants were between 19 to 28 ages old. There were five females, 14 males and one non-binary. Eleven participants reported touching their face at least once per hour, six at least once per minute, one at least once per day and two unaware how often they touched their face. Participants were also asked to describe five or more different ways that they would {\it normally} touch their face in their everyday lives, the result of which we show in \fgref{survey}.
As in any other survey, asking participants to estimate frequency of an activity subjects to inaccuracy of the memory. However, our goal is not to compute an exact frequency but to elicit a list of commonly-occurring face touching behaviors.

\section{Model Hyperparameter Tuning}
We implemented Random Forest Models using the scikit-learn Python library\footnote{\url{https://scikit-learn.org/}}.
To further improve the model, we used the training data set to perform a hyperparameter tuning by first performing a randomized search to narrow down to ranges of parameters, within which we then employed a grid search to pinpoint specific optimal parameters. 
We repeated this process five times to obtain five sets of parameters for models corresponding to $T = \{1.0s, 1.1s, 1.2s, 1.3s, 1.4s, 1.5s\}$, as shown below. We use scikit-learn's default values for the rest of the parameters.

\begin{table}[H]
    \small
    \begin{tabular}{lcccccc}
    \toprule
    t~=                 & 1.0 & 1.1   & 1.2   & 1.3   & 1.4   & 1.5   \\
    \midrule
    bootstrap           & False & False & False & False & False & False \\
    max\_depth          & 200   & 150 & 150   & 300   & 200   & 150   \\
    max\_features       & log2  & log2 & log2  & log2  & log2  & log2  \\
    min\_samples\_leaf  & 2     & 1 & 1     & 1     & 4     & 2     \\
    min\_samples\_split & 3     & 2 & 2     & 3     & 4     & 3     \\
    n\_estimators       & 150   & 150 & 300   & 200   & 100   & 300  \\
    \bottomrule
\end{tabular}
\end{table}

\section{Open Science}
The training/testing datasets, model and source code are available at {\url https://hci.ucla.edu/faceoff} to spur the future development of face touching detection to combat the COVID-19 pandemic.

\end{document}